\documentclass[RNAAS]{aastex63}

\begin{document}

\title{Satellite Constellation Internet Affordability and Need}

\correspondingauthor{Meredith L. Rawls}
\email{mrawls@uw.edu}

\author[0000-0003-1305-7308]{Meredith L. Rawls}
\affiliation{Department of Astronomy/DIRAC Institute, University of Washington, Seattle, WA, USA}

\author[0000-0003-4195-5068]{Heidi B. Thiemann}
\affiliation{School of Physical Sciences, The Open University, Milton Keynes, MK7 6AA, UK}

\author[0000-0001-7286-3401]{Victor Chemin}
\affiliation{Universit\'{e} de Rouen Normandie, Rouen, France}

\author[0000-0003-2918-8687]{Lucianne Walkowicz}
\affiliation{The Adler Planetarium/The JustSpace Alliance, Chicago, IL, USA}

\author[0000-0003-3412-2586]{Mike W. Peel}
\affiliation{Instituto de Astrof\'{i}sica de Canarias, 38200 La Laguna, Tenerife, Canary Islands, Spain}
\affiliation{Departamento de Astrof\'{i}sica, Universidad de La Laguna (ULL), 38206 La Laguna, Tenerife, Spain}

\author[0000-0001-5125-9539]{Yan G. Grange}
\affiliation{ASTRON, Netherlands Institute for Radio Astronomy, Oude Hoogeveensedijk 4, NL-7991 PD Dwingeloo, the Netherlands}

\received{12 October 2020}
\accepted{23 October 2020}
\published{27 October 2020}
%\submitjournal{RNAAS}

%\keywords{artificial satellites (68)}

\begin{abstract}
    Large satellite constellations in low-Earth orbit seek to be the infrastructure for global broadband Internet and other telecommunication needs. We briefly review the impacts of satellite constellations on astronomy and show that the Internet service offered by these satellites will primarily target populations where it is unaffordable, not needed, or both. The harm done by tens to hundreds of thousands of low-Earth orbit satellites to astronomy, stargazers worldwide, and the environment is not acceptable.
    
    \vspace{0.5em}
    \noindent\textit{Unified Astronomy Thesaurus concepts:} \href{http://vocabs.ands.org.au/repository/api/lda/aas/the-unified-astronomy-thesaurus/current/resource.html?uri=http://astrothesaurus.org/uat/68}{Artificial satellites (68)}
\end{abstract}

%-------------------------------------
\section{Low-Earth orbit satellite constellations}

SpaceX's Starlink has launched over 700 low-Earth orbit satellites since May 2019. They plan to offer global broadband Internet with a final ``constellation'' of 42,000 satellites. Other operators
%including Amazon Kuiper and OneWeb
have announced similar plans,
%\footnote{\url{https://www.linkedin.com/pulse/new-fcc-processing-round-prompts-license-requests-frhr-von-der-ropp/}}
and we are witnessing a new era of a sky filled with thousands of low-Earth orbit commercial satellites.

Recent studies (\citealt{2020ApJ...892L..36M,2020A&A...636A.121H,2020A&A...637L...1T})
%as well as presentations and commentaries (\citealt{2020AAS...23631107S}\footnote{\url{https://nsf.gov/attachments/299316/public/12_Satellite_Constellations_and_Astronomy-Pat_Seitzer.pdf}}; \citealt{2020Mirro...1...63T,rawls_meredith_2020_3937869,Tyson1543})
raise concerns about the brightness of these constellations and the detrimental effect of large numbers of
%low-Earth orbit
satellites to optical astronomy. This is because satellites reflect sunlight even after sunset, and global satellite coverage will cause bright streaks in astronomical images for large portions of the night at ground-based observatories worldwide.

Vera C. Rubin Observatory and its Legacy Survey of Space and Time \citep[LSST,][]{2019ApJ...873..111I}
%a top priority in the 2010 Decadal Survey of Astronomy \& Astrophysics \citep{NAP12951},
%will see first light in the early 2020s.%\footnote{\url{https://docushare.lsst.org/docushare/dsweb/Get/Document-35216}}.
%Rubin Observatory
will be the optical astronomy facility most severely impacted by satellite constellations due to its wide field of view and large light collecting area. \citet{2020arXiv200612417T} estimate a 48,000 satellite constellation will result in at least 30\% of LSST images containing a satellite trail.
Other optical and near-IR observatories will also be significantly impacted.
There are concerns for wavelengths outside optical (e.g.,~\citealt{2020arXiv200305472G,Massey}) as well. 
%Radio astronomy will be particularly affected
For example, satellites directly transmit at 10--30\,GHz\footnote{\url{https://docs.fcc.gov/public/attachments/FCC-18-38A1.pdf}} in bands used for astronomical observations.
Depending on satellite transmitter quality, frequencies outside the nominal transmission bands can also be impacted \citep[e.g.,][]{2019JAI.....840009D}.

Unfortunately, the solution is not as simple as moving telescopes into space. Prohibitive costs, inability to maintain instruments in space, constraints imposed by launch vehicle size, the short
%$\sim$5-year
life expectancy from harsh conditions, and decades to plan and fund large missions make this infeasible.
%Space telescopes are also not immune to the effects of satellite constellations.
%The infusion of thousands of satellites into the already-crowded orbital environment\footnote{\url{https://www.newyorker.com/magazine/2020/09/28/the-elusive-peril-of-space-junk}} has caused an increase in collision avoidance maneuvers \citep{Hongqiang_2020}.

The recent Satellite Constellations 1 Workshop Report \citep{satcon1} provides recommendations
%for astronomers and satellite operators
to mitigate the effects of satellite trails in optical and near-IR images. Final reports from the Conference on Dark and Quiet Skies for Science and Society\footnote{\url{http://research.iac.es/congreso/quietdarksky2020/pages/home.php}} will soon be sent to the UN Committee on the Peaceful Uses of Outer Space.
%Any resulting regulations would be in addition to existing rules such as the Outer Space Treaty\footnote{\url{https://www.unoosa.org/oosa/en/ourwork/spacelaw/treaties/outerspacetreaty.html}}.

SpaceX is collaborating with Rubin Observatory, and
%representatives from 
other satellite operators
including Amazon Kuiper and OneWeb
have begun dialogues with astronomers too. While early Starlink mitigations are promising, it is unrealistic to rely on the goodwill of satellite operators as a mitigation strategy. Hardware and software mitigations can minimize some scientific impacts, but they are a significant amount of work that has not been planned or budgeted for. They also do not address wider-reaching environmental or cultural effects of a drastically changed night sky.

%-------------------------------------
\section{Affordable Internet access?}

\begin{figure}[ht!]
    \centering
    \includegraphics[width=\textwidth]{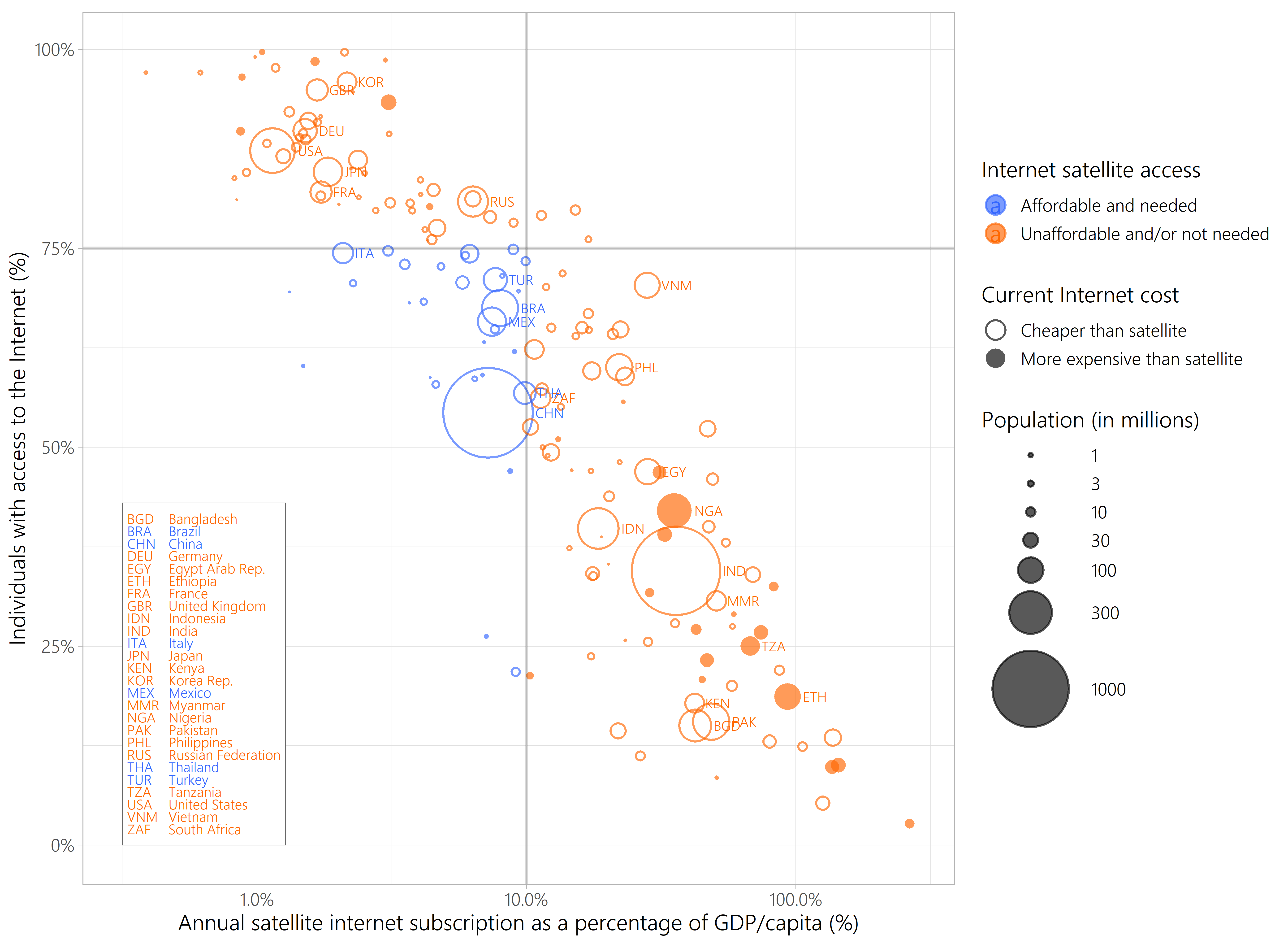}
    \caption{There is a small population that both needs and can afford satellite Internet service (blue circles). The remainder either does not need or cannot afford it (orange circles). For each country, we plot the estimated cost of satellite constellation Internet service as a percentage of gross domestic product (GDP) per capita against percentage of individuals currently with Internet access. Open circles indicate current ground-based Internet is cheaper than estimated satellite Internet. Circle size indicates the country's population. Affordable access is classed as $\leq$10\% of GDP per capita; need is classed as $\leq$75\% of the population currently having Internet access. Countries with $>$50 million inhabitants are labelled, accounting for 80\% of the world population. Data and code at \url{https://github.com/mrawls/sky-high-starlink} \& \url{https://doi.org/10.5281/zenodo.4133883}.
    }
    \label{fig:thefig}
\end{figure}

While large low-Earth orbit satellite constellations will harm astronomy and may render certain orbits hazardous
% due to Kessler syndrome
\citep{kessler2010,Hongqiang_2020}, such negative impacts could be considered acceptable if constellations offer substantial benefits.
The rationale often given is a strong need for affordable Internet access worldwide. This need certainly exists, and SpaceX is currently offering free beta Starlink Internet to communities in need including emergency first responders
%\footnote{\url{https://www.theverge.com/2020/9/29/21493158/spacex-starlink-Internet-washington-emergency-wildfires-first-responders}} 
and the Hoh Tribe
%\footnote{\url{https://www.pcmag.com/news/native-american-tribe-gets-early-access-to-spacexs-starlink-and-says-its}} 
in Washington. However, providing free Internet access to communities in need is not Starlink's primary goal, nor is it a sustainable business model.
To the contrary, we show in Figure \ref{fig:thefig} that the cost of a satellite Internet subscription remains out of reach to the communities that need it most.

In Figure \ref{fig:thefig}, we use Starlink as an example, but emphasize the calculation is similar for other satellite Internet providers.
We parameterize affordability as the estimated annual cost of a satellite Internet subscription as a percentage of gross domestic product (GDP) per capita.
We adopt a conservative estimate of \$60 monthly for satellite Internet service. This is lower than the \$80 monthly mentioned by SpaceX President Gwynne Shotwell\footnote{\url{https://edition.cnn.com/2019/10/26/tech/spacex-starlink-elon-musk-tweet-gwynne-shotwell/index.html}} and in line with an internal note stating SpaceX plans to make \$40 billion annually with 30 million subscribers in 2025\footnote{\url{https://www.wsj.com/articles/exclusive-peek-at-spacex-data-shows-loss-in-2015-heavy-expectations-for-nascent-Internet-service-1484316455}}. We use Internet access and GDP data from 2018, which are publicly available from the UN International Telecommunication Union
(ITU), the UN specialized agency for information and communication technologies\footnote{\url{https://www.itu.int/net4/ITU-D/icteye/}},
and the World Bank\footnote{\url{https://data.worldbank.org/indicator/NY.GDP.PCAP.CD?end=2019&start=1960}}. 

The UN Sustainable Development Goal 9 (SDG9) is ``Investing in Information and Communication Technologies (ICT) access and quality education to promote lasting peace'' \citep{/content/publication/55eb9109-en}.
In developing countries, fewer than 50\% of people have access to the Internet, and in the least developed countries, it falls to fewer than 20\%\footnote{\url{https://www.itu.int/en/ITU-D/Statistics/Documents/facts/FactsFigures2019.pdf}}.
To satisfy SDG9, there is a need for massive investment in ICT in
remote and vulnerable communities in
developing countries as well as improved Internet access in developed rural areas\footnote{\url{https://blogs.microsoft.com/on-the-issues/2019/04/08/its-time-for-a-new-approach-for-mapping-broadband-data-to-better-serve-americans/}}.
SpaceX aims to provide low-latency Internet to 80--100 million households, or 3--4\% of the world's population\footnote{\url{https://www.youtube.com/watch?v=HPV8Xp3pEpI}}, with a bandwidth suitable only for low- to mid-density populations\footnote{\url{https://arstechnica.com/information-technology/2020/03/musk-says-starlink-isnt-for-big-cities-wont-be-huge-threat-to-telcos/}}. Technological constraints suggest other satellite operators can offer a similar service.

Given this, it would appear satellite constellation Internet can contribute to SDG9, particularly in the ``blue'' countries in Figure \ref{fig:thefig}. However, the estimated \$80 monthly cost of a Starlink subscription is prohibitive to all countries in greatest need of access, and this does not include start-up costs of \$100--300\footnote{\url{https://www.youtube.com/watch?t=12m37s&v=AHeZHyOnsm4}} per user.
In countries where the majority of the population lives on less than a few dollars per day, the ability to spend \$80 per month for Internet access is infeasible.
Even if subscriptions are heavily subsidised (e.g., \$20-30 per month), it still remains inaccessible for many. Additionally, some countries are already working to provide cheaper ground-based broadband. For example, Nigeria
%aims to reach 70\% coverage by 2025 and
already has 4G mobile Internet for \$27 monthly\footnote{\url{https://africanbusinessmagazine.com/sectors/technology/nigeria-rolls-out-broadband-to-boost-growth/}}.
For the few people that both need and can afford satellite Internet, the majority reside in China, a country currently developing their own constellations\footnote{\url{http://www.circleid.com/posts/20201002-a-new-chinese-broadband-satellite-constellation/}}.
% and unlikely to subscribe to Starlink.

%-------------------------------------
\section{Questioning the need for satellite constellations}

The choice between Internet access and astronomy is not a binary one, but we are racing toward a tipping point of no return and a future with tens to hundreds of thousands of satellites. 
While there undeniably is a need for high-speed, accessible, affordable Internet access, corporate satellite constellations are not humanitarian projects that freely provide infrastructure for SDG9.
Figure \ref{fig:thefig} demonstrates that satellite constellations will not on the whole provide Internet to those who need it most. The rush to launch tens of thousands of satellites should be fundamentally reconsidered.

We use Starlink as an example
in this Note
because SpaceX is the first company to launch hundreds of satellites and engage in discussion with astronomers. This collaboration is likely the best-case scenario, since other operators currently have no financial or regulatory motive to ensure they darken or otherwise mitigate the effects of their satellites. We recognize that satellite constellations are accepted by many as the new norm, and that
simply
not launching them may be perceived as unrealistic. We emphasize that we are not advocating for a world with zero satellites, and we sincerely appreciate continued dialogues with
satellite
operators to implement a variety of technical mitigations.
At the same time, the night sky is an invaluable resource that must not be exploited for profit. It must be protected, not just for the freedom of scientific exploration and cultural heritage now, but for future generations.

\acknowledgements

The earliest form of Figure \ref{fig:thefig} was created by Victor Chemin and shared on Twitter. Lucianne Walkowicz presented it in a talk at DotDotAstro (\url{https://www.dotastronomy.com/alpha}), after which this Note was drafted.

\newpage

\bibliography{bibliography}{}

\begin{thebibliography}{}
\expandafter\ifx\csname natexlab\endcsname\relax\def\natexlab#1{#1}\fi
\providecommand{\url}[1]{\href{#1}{#1}}
\providecommand{\dodoi}[1]{doi:~\href{http://doi.org/#1}{\nolinkurl{#1}}}
\providecommand{\doeprint}[1]{\href{http://ascl.net/#1}{\nolinkurl{http://ascl.net/#1}}}
\providecommand{\doarXiv}[1]{\href{https://arxiv.org/abs/#1}{\nolinkurl{https://arxiv.org/abs/#1}}}

\bibitem[{{Deshpande} \& {Lewis}(2019)}]{2019JAI.....840009D}
{Deshpande}, A.~A., \& {Lewis}, B.~M. 2019, Journal of Astronomical
  Instrumentation, 8, 1940009, \dodoi{10.1142/S2251171719400099}

\bibitem[{{Gallozzi} {et~al.}(2020){Gallozzi}, {Paris}, {Scardia}, \&
  {Dubois}}]{2020arXiv200305472G}
{Gallozzi}, S., {Paris}, D., {Scardia}, M., \& {Dubois}, D. 2020, arXiv
  e-prints, arXiv:2003.05472.
\newblock \doarXiv{2003.05472}

\bibitem[{{Hainaut} \& {Williams}(2020)}]{2020A&A...636A.121H}
{Hainaut}, O.~R., \& {Williams}, A.~P. 2020, \aap, 636, A121,
  \dodoi{10.1051/0004-6361/202037501}

\bibitem[{Hongqiang \& Zhanyue(2020)}]{Hongqiang_2020}
Hongqiang, S., \& Zhanyue, Z. 2020, {IOP} Conference Series: Earth and
  Environmental Science, 552, 012014, \dodoi{10.1088/1755-1315/552/1/012014}

\bibitem[{{Ivezi{\'c}} {et~al.}(2019){Ivezi{\'c}}, {Kahn}, {Tyson},
  {et~al.}}]{2019ApJ...873..111I}
{Ivezi{\'c}}, {\v{Z}}., {Kahn}, S.~M., {Tyson}, J.~A., {et~al.} 2019, \apj,
  873, 111, \dodoi{10.3847/1538-4357/ab042c}

\bibitem[{Kessler {et~al.}(2010)Kessler, Johnson, Liou, \&
  Matney}]{kessler2010}
Kessler, D., Johnson, N., Liou, J.-C., \& Matney, M. 2010, Advances in the
  Astronautical Sciences, 137

\bibitem[{{Massey}(2020)}]{Massey}
{Massey}, R. 2020, Astronomy and Geophysics, 61, 2.19,
  \dodoi{10.1093/astrogeo/ataa027}

\bibitem[{{McDowell}(2020)}]{2020ApJ...892L..36M}
{McDowell}, J.~C. 2020, \apjl, 892, L36, \dodoi{10.3847/2041-8213/ab8016}

\bibitem[{{Tregloan-Reed} {et~al.}(2020){Tregloan-Reed}, {Otarola}, {Ortiz},
  {Molina}, {Anais}, {Gonz{\'a}lez}, {Colque}, \&
  {Unda-Sanzana}}]{2020A&A...637L...1T}
{Tregloan-Reed}, J., {Otarola}, A., {Ortiz}, E., {et~al.} 2020, \aap, 637, L1,
  \dodoi{10.1051/0004-6361/202037958}

\bibitem[{{Tyson} {et~al.}(2020){Tyson}, {Ivezi{\'c}}, {Bradshaw}, {Rawls},
  {Xin}, {Yoachim}, {Parejko}, {Greene}, {Sholl}, {Abbott}, \&
  {Polin}}]{2020arXiv200612417T}
{Tyson}, J.~A., {Ivezi{\'c}}, {\v{Z}}., {Bradshaw}, A., {et~al.} 2020, arXiv
  e-prints, arXiv:2006.12417.
\newblock \doarXiv{2006.12417}

\bibitem[{{United Nations}(2019)}]{/content/publication/55eb9109-en}
{United Nations}. 2019, The Sustainable Development Goals Report 2019 (UN), 61.
\newblock \url{https://www.un-ilibrary.org/content/publication/55eb9109-en}

\bibitem[{{Walker} {et~al.}(2020){Walker}, {Hall}, {Allen}, {Green}, {Seitzer},
  {Tyson}, {Bauer}, {Krafton}, {Lowenthal}, {Parriott}, {Puxley}, {Abbott},
  {Bakos}, {Barentine}, {Bassa}, {Blakeslee}, {Bradshaw}, {Cooke}, {Devost},
  {Galad{\'\i}-Enr{\'\i}quez}, {Haase}, {Hainaut}, {Heathcote}, {Jah},
  {Krantz}, {Kucharski}, {McDowell}, {Mr{\'o}z}, {Otarola}, {Pearce}, {Rawls},
  {Saunders}, {Seaman}, {Siminski}, {Snyder}, {Storrie-Lombardi},
  {Tregloan-Reed}, {Wainscoat}, {Williams}, \& {Yoachim}}]{satcon1}
{Walker}, C., {Hall}, J., {Allen}, L., {et~al.} 2020, in Bulletin of the
  American Astronomical Society, Vol.~52, 0206,
  \dodoi{10.3847/25c2cfeb.346793b8}

\end{thebibliography}
\bibliographystyle{aasjournal}

\end{document}